\documentclass{ws-ijmpa}
\usepackage{graphicx}
\usepackage{color}
\usepackage[latin1]{inputenc}
\newcommand{\be}{\begin{eqnarray}}
\newcommand{\ee}{\end{eqnarray}}
\newcommand{\nn}{~\nonumber \\}

\newcommand{\bmp}{\noindent\begin{minipage}{16cm}}
\newcommand{\emp}{\end{minipage}\vskip 7mm} 

\usepackage{graphicx}
\usepackage{dcolumn}
\usepackage{bm}
\usepackage{amsmath}
\usepackage{amsfonts}
\usepackage{bbm}
\usepackage{subfigure} 

\def\drawbox#1#2{\hrule height#2pt
        \hbox{\vrule width#2pt height#1pt \kern#1pt
              \vrule width#2pt}
              \hrule height#2pt}

\def\Asym#1#2{\vcenter{\vbox{\drawbox{#1}{#2}
              \kern-#2pt 
              \drawbox{#1}{#2}}}}

\begin{document}

\markboth{Francesco Sannino}
{{ L}ight {C}omposite { H}iggs: LCH @ LHC }

%
\catchline{}{}{}{}{}
%

\title{Light Composite {H}iggs: LCH @ LHC}

\author{Francesco Sannino}

\address{The Niels Bohr Institute, Blegdamsvej 17,\\
Copenhagen \O, DK-2100,Denmark.\\
sannino@nbi.dk}

\maketitle

\begin{abstract}
Here I summarize some of the salient features of technicolor theories with technifermions in higher dimensional representations of the technicolor gauge group. The expected phase diagram as function of number of flavors and colors for the two index (anti)symmetric representation of the gauge group is reviewed. After having constructed the simplest walking technicolor theory one can show that it is not at odds with the precision measurements. The simplest theory also requires, for consistency, a fourth family of heavy leptons. The latter may result in an interesting signature at LHC. In the case of a fourth family of leptons with ordinary lepton hypercharge the new heavy neutrino can be a natural candidate of cold dark matter. New theories will also be proposed in which the critical number of flavors needed to enter the conformal window is higher than in the one with fermions in the two-index symmetric representation, but lower than in the walking technicolor theories with fermions only in the fundamental representation of the gauge group. Due to the near conformal/chiral phase transition the composite Higgs is very light compared to the intrinsic scale of the technicolor theory. For the two technicolor theory the composite Higgs mass is predicted not to exceed 150 GeV
\keywords{Electroweak Symmetry Breaking; Strong Dynamics; Technicolor.}
\end{abstract}

\ccode{PACS numbers: 11.25.Hf, 123.1K}

\section{Introduction}
Determining the nature of the Higgs boson is one of the most important problems of theoretical physics. The Large Hadron Collider experiment (LHC) at CERN will be soon shedding light on this sector of the electroweak theory.

Of particular interest to us are models of electroweak symmetry breaking via new strongly interacting theories of technicolor type \cite{TC}. This is a mature subject (for recent reviews see \cite{Hill:2002ap,{Lane:2002wv}}) where considerable 
effort has been made to construct viable models. One of the main difficulties in constructing such extension of the standard model is the very limited knowledge about generic strongly interacting theories. This has led theorists to consider specific models of technicolor which resemble ordinary quantum chromodynamics and for which the large body of experimental data at low energies can be directly exported to make predictions at high energies. According to Peskin and Wells \cite{Peskin:2001rw} generic theories of composite Higgs contain large corrections with respect to the minimal standard model, similar to those of a heavy elementary Higgs boson \cite{Peskin:1991sw}. However, a heavy composite Higgs is not always 
an outcome of strong dynamics \cite{{Sannino:2004qp},Hong:2004td,{Dietrich:2005jn}}. Here we are not referring to models in which the Higgs is a quasi Goldstone boson \cite{Dimopoulos:1981xc} which have been 
investigated recently \cite{Arkani-Hamed:2001nc}.

{ Some of the problems of the simplest technicolor models, such as
providing ordinary fermions with a mass, are
alleviated when considering new gauge dynamics in which the
coupling does not run with the scale but rather walks, i.e.
evolves very slowly \cite{Holdom:1981rm,Yamawaki:1985zg,Appelquist:an,MY,{Frere:1986ct}} (i.e. walking technicolor).
Most of the investigations in the literature used matter in the fundamental representation of the gauge group. In this case one needs a very large number of matter fields, roughly of the order of $4N$ with $N$ the number of technicolors  to achieve the walking. A large number of techniflavors has many shortcomings, such 
as large contributions to the oblique parameters and a very large number of unwanted Goldstone bosons. 
However it was shown recently \cite{Sannino:2004qp,{Hong:2004td}} that it is possible to consider matter 
in higher dimensional representations and achieve walking for a very small number of fields. 

A simple way to understand why the present theories are still viable is that the resulting composite Higgs is lighter than the one typically expected for conventional composite Higgs theories. Remarkably, the theories investigated here not only explain the hierarchy problem but also lead to a light composite Higgs.  We note that technicolor like theories with fermions in higher dimensional representation of the gauge group have been considered in the literature
\cite{Lane:1989ej,{Lane:1991qh},{Corrigan:1979xf}}. 

We also propose theories in which the critical number of flavors needed to enter the conformal window is higher than the one with 
 fermions in the two-index symmetric representation, but lower than the traditional walking technicolor theories with fermions 
 in the fundamental representation of the gauge group. A simple class of these theories are split (super)technicolor theories in which 
 we add only a (techni)gluino to the theory with still $N_{Tf}$ fermions in the fundamental representation of the gauge group.
 

\section{Features of Higher Representations} 
\label{2}

\subsection{The Phase Diagram}

Here I summarize some of the basic features of the theories with two-index representations 
explored in \cite{Sannino:2004qp,{Hong:2004td},{Dietrich:2005jn}}. These theories have
fermions in the two-index symmetric (S-type) or antisymmetric
(A-type) representation. 


The relevant feature, discovered in \cite{Sannino:2004qp}, is that the
S-type theories can be near conformal already at $N_{T f}=2$ when $N=2$ or $3$. This
should be contrasted with theories in which the fermions are in the fundamental
representation for which the minimum number of flavors required to
reach the conformal window is eight for $N=2$.

The $N=3$ model with A-type fermions is just $N_{T f}$-flavor QCD with the maximum allowed number of flavors equal to $16$. For $N=2$
the antisymmetric representation goes over to a pure Yang-Mills theory
with a singlet fermion. For S-type models, asymptotic freedom is
lost already for three flavors when $N=2$ or $3$, while the upper
bound of $N_{T f}=5$ is reached for $N=20$ and does not change when
$N$ is increased further. The phase diagram, studied in \cite{Sannino:2004qp} as a function of the number of colors and flavors
for the S- and A-type theories is summarized in figure \ref{PhaseDiagram}.
\begin{figure}[t]

\begin{center}\includegraphics[width=10truecm,height=3truecm]{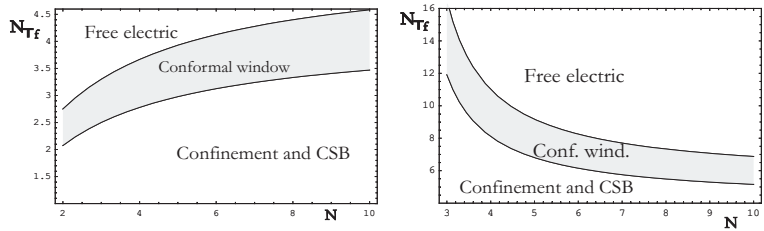}
\end{center}
\caption{Left(Right) panel: Phase diagram as function of number of $N_{Tf}$ 
Dirac
flavors and $N$ colors for fermions in the two-index symmetric (antisymmetric)
representation, i.e. S(A)-types, of the gauge group.} \label{PhaseDiagram}
\end{figure}

{}From the picture it is clear that for $N=2,3,4,5$ $N_{Tf}=2$ is already the highest
number of flavors possible before entering the conformal window.
Hence, for these theories we expect a slowly evolving coupling
constant. The critical number of flavors must be greater than three for $N\geq 6$, but
remains smaller than or equal to four for any $N$.

The critical value of flavors increases with the number of colors
for the gauge theory with S-type matter: the limiting value is
$4.15$ at large $N$. 
These estimates are based on the validity of the first
few terms in the perturbative expansion of the $\beta$-function.

The situation is different for the theory with A-type matter. As it is evident
from the phase diagram,
the critical number of flavors increases when decreasing the
number of colors making them phenomenologically inadequate.

\subsection{Possible alternatives: Split Technicolor}
If we insist on keeping the technifermions in the fundamental representation while trying to reduce the number of 
techniflavors needed to be near a conformal window another possibility is to add matter uncharged under the weak interactions. In this 
way one would, in general, increase the number of pseudogolstone bosons. However an interesting and minimal possibility is to consider 
adding a massless Weyl fermion in the adjoint representation of the gauge group. This particle behaves as a technigluino. The 
resulting theory has the same matter content as $N_{Tf}$-flavor super QCD but without the scalars. This split technicolor theory 
has the critical number of flavors above which one enters the conformal window lying within the range:
\begin{eqnarray} 
\frac{3}{2}< \frac{N^{c}_{Tf}}{N} \lesssim 4 \ .
\end{eqnarray}
Here $N$ is the number of colors. The lower bound is the exact supersymmetric point for a non-perturbative 
conformal fixed point \cite{Intriligator:1995au} while the upper 
bound is the one expected in the theory without a technigluino \cite{Dietrich:2005jn}. Interestingly, this shows that with two colors the number of (techni)flavors needed to be near 
the conformal window in the split case is at least three, while for three colors more than five flavors are required. 
These values are still larger than the ones presented 
for theories with fermions in the two-index symmetric representation, although still lower than the ones used in 
walking technicolor theories with fermions only in the fundamental representation of the gauge group. It is 
useful to remind the reader that in supersymmetric theories the critical number of flavors needed to enter the 
conformal window does not coincide with the critical number of flavors required to restore chiral symmetry. The 
scalars in supersymmetric theories play an important role from this point of view. 
We note that a split technicolor-like theory has been used recently in \cite{Hsu:2004mf}, to investigate the 
strong CP problem. 

Split technicolor is a theory sharing some features with theories of split supersymmetry recently advocated and 
studied in \cite{Arkani-Hamed:2004fb} as possible extensions of the 
standard model. Clearly we have introduced split technicolor, differently from split supersymmetry, to address 
the hierarchy problem. We, however, focus on theories with 
fermions in higher representation of the gauge group.
\section{Two Technicolors and two flavor and a New Lepton Family}
\label{3}

Here, the gauge dynamics driving electroweak symmetry breaking consists
of two Dirac fermions in the two-index symmetric representation
of the $SU(2)$ gauge theory. 
This model is the one with the smallest
perturbative $S$-parameter \cite{Hong:2004td} which preserves the following
relevant feature: It is (quasi)conformal with {\it just} one doublet of
techinfermions \cite{Sannino:2004qp}. This naturally leads to a two
scale theory \cite{Cohen:1988sq}: The lowest scale is the one at which the
coupling constant becomes strong (i.e. the electroweak scale). The
other scale is defined to be the one above which the $SU(2)$ of technicolor gauge
coupling constant starts running. These two scales are exponentially
separated. 
This fact allows us to concentrate on the physics near
the electroweak symmetry breaking scale. Using a bottom-up type of approach we postpone questions associated to the detailed
dynamics of the generation of fermion masses.

We represent the doublet of techni-fermions as:
\be
T_L^{\{C_1,C_2 \}}
=
\left(\begin{array}{l}U^{\{C_1,C_2 \}}\\ D^{\{C_1,C_2 \}}\end{array}\right)_L \ ,
\qquad 
T_R^{\{C_1,C_2\}}&=&\left(U_R^{\{C_1,C_2\}},~ D_R^{\{C_1,C_2\}}\right) \ .
\nn
\ee
Here $C_i=1,2$ is the technicolor index and $T_{L(R)}$ is a doublet (singlet) with respect 
to the weak interactions. 
The two-index symmetric representation of $SU(2)$ is real, and hence the global classical
symmetry group is $SU(4)$ which breaks to $O(4)$. This leads to the appearance of nine
Goldstone bosons, of which three become the longitudinal components of the weak gauge bosons.
The low energy spectrum is expected
to contain six quasi Goldstone bosons which receive mass through
extended technicolor (ETC) interactions \cite{Hill:2002ap,{Lane:2002wv},Appelquist:2002me,Appelquist:2003uu,{Appelquist:2004es},{Appelquist:2004ai},{Appelquist:2004mn}}.

As pointed out in \cite{Sannino:2004qp}, the weak interactions are also affected by the $SU(2)$ Witten anomaly
\cite{Witten:fp}. More specifically, since our techniquarks are in the two-index symmetric 
representation of $SU(2)$ we have exactly three extra left doublets from the point of view of the weak interactions. 
A simple way to cure such an anomaly without introducing further unwanted
gauge anomalies is to introduce at least one new lepton family. According to the choice of the hypercharge we discuss 
a number of relevant cases:

\subsection{New Standard Model Like Lepton Family}
Since we have three doublets of techniquarks which resemble very much an ordinary triplet of colored quarks one can assign to the 
techniquarks 
the standard quark hypercharge which for the left-handed technifermions is then $Y=+1/6$.
The hypercharge is linked to the ordinary charge following the convention 
$Q = T_3 + {Y}$.
 This yields:\be
T_L^{( Q)}
=
\left(\begin{array}{l} U^{(+2/3)}\\D^{(-1/3)}\end{array}\right)_L
\ee
where we have provided the electric charges of the techniquarks and suppressed the technicolor indices. {}For the right-handed techni-fermions which are isospin singlets we have:
\be
T_R^{(Q)}=\left( U_R^{(+2/3)},~ D_R^{(-1/3)}\right) \ ,
\qquad 
Y=+\frac{2}{3},~~~~-\frac{1}{3} \ .
\ee
In this case it is sufficient to add one new generation of left-handed leptons with hypercharge $Y=-1/2$:
\be
{\cal L}_L ^{(Q)}=
\left(\begin{array}{l}\nu_{\zeta}^{(0)} \\ \zeta^{(-1)}\end{array}\right)_L \ .
\ee
Clearly this new lepton family must be sufficiently heavy and not at odds with the electroweak precision measurements. 

In the case of a pure Dirac mass term we add one generation of right-handed leptons (isospin singlets)
\be
{\cal L}_R^{(Q)}&=&\left({\nu_{\zeta}}_R^{(0)},~\zeta_R^{(-1)}\right) \ ,
\qquad 
Y=0,~~-1 \ .
\ee
If the neutral lepton is lighter than the associated charged lepton and since, by assumption, it does not mix with the lighter lepton generations, it becomes absolutely stable and a potential natural candidate for cold dark matter. This is so since a fourth stable neutrino has only weak interactions. Besides, we note that since the fourth family of leptons in our scenario is needed to compensate for the anomalies introduced by the techniquarks, its mass scale is naturally linked to the electroweak symmetry breaking scale. There are, however, a few caveats which need to be resolved to make the present neutrino a reasonable candidate for cold dark matter. Indeed if it turns out to be absolutely stable it will have to cluster \cite{Ringwald:2004np} rather than be distributed homogeneously in the universe as is assumed when deriving bounds on its mass \cite{reusser91,{Abusaidi:2000wg}}. In any case this neutrino can be 
a relevant component of cold dark-matter. Precision electroweak measurements, as we shall see, are already able to provide relevant information on the neutral versus charged lepton mass. 

{}From a theoretical point of view one can also provide a Majorana type mass for the new neutrino. In this case we need not to introduce the right neutrino field and the fourth family continuous lepton number is not conserved. The residual possible $Z_2$ symmetry under which only the neutrino transforms as
\begin{eqnarray}
{\nu_{\zeta}}_L \rightarrow -{\nu_{\zeta}}_L \ , \qquad \zeta_L \rightarrow \zeta_L \ ,
\end{eqnarray}
although left unbroken by the mass term, is violated by the weak interactions. {}For this type of heavy Majorana neutrino the bounds derived in \cite{reusser91} are much weaker even when assuming this type of matter does not cluster. It might then be a better candidate for cold dark matter. In \cite{Dietrich:2005jn} we have also considered the case of both, a Dirac and a Majorana mass.

\subsection{A more general  hypercharge assignment}
In general, all of the anomalies are avoided using the following 
generic hypercharge assignment:
\begin{eqnarray}
Y(T_L)& = &\frac{y}{2} \ ,\qquad Y(U_R,D_R)=\left(\frac{y+1}{2},\frac{y-1}{2}\right) \ , \\
Y({\cal L}_L)& = &-3\frac{y}{2} \ ,\qquad Y({\nu_{\zeta}}_R,\zeta_R)=\left(\frac{-3y+1}{2},\frac{-3y-1}{2}\right) \ .
\end{eqnarray}
One recovers the previous choices of the hypercharge for $y=1/3$ (standard model like family) and $y=0$ (fractionally charged leptons). Another choice of the hypercharge which does not lead to either fractionally charged techniquarks or leptons is, for example, $y=1$. In this case:
\begin{eqnarray}
Q(U)=1 \ , \quad Q(D)=0 \ , \quad Q(\nu_{\zeta})=-1 \ , \quad {\rm and }\quad Q(\zeta)=-2 \ , \quad {\rm with} \quad y=1 
\end{eqnarray}
We will also analyze this possibility from the point of view of the electroweak precision measurements. In this case the dark matter candidate could be a neutral ditechniquark with the quantum numbers, for example, of 
two $D$ techniquarks. A more complete analysis is however needed along the lines suggested first in \cite{Chivukula:1989qb} for ordinary technicolor theories.

For a three technicolor theory we refer the reader to the discussion in \cite{Dietrich:2005jn}.  
\section{General Constraints from Electroweak Precision Data}
\label{5}

In this section we confront our models with the electroweak precision measurements. The relevant corrections to the minimal standard model appear in the vacuum polarizations of the electroweak gauge bosons. These can be parameterized in terms of the three 
quantities $S$, $T$, and $U$ (the oblique parameters) 
\cite{Peskin:1990zt,Peskin:1991sw,Kennedy:1990ib,Altarelli:1990zd}, and confronted with the electroweak precision data. The relevant formulae we use are from \cite{He:2001tp}.
The models considered here produce smaller values of $S$ than
traditional technicolor models, because of their smaller particle
content and because of their near-conformal dynamics \cite{Hong:2004td}.

The simple perturbative estimate leads to the following value of the $S$ parameter:
\begin{equation}
{ S}_{\rm pert.} = {1 \over 2 \pi}  \ .
\end{equation}
 This is the walking technicolor theory with the smallest perturbative $S$ parameter \cite{Hong:2004td}. 
However one cannot compare this result immediately with the electroweak precision data since this theory has a Witten anomaly and hence the new lepton family must be included in the analysis when comparing with the precision measurements.

Here we consider the combined effect on the electroweak precision measurements of this walking technicolor theory 
together with the fourth family of leptons. If we choose the hypercharge for the leptons to be the one of ordinary leptons one can easily 
see that from the electroweak point of view the theory features effectively a complete fourth family of leptons
 and (techni)quarks. This is so since the techniquark doublet, being in the two-index symmetric representation of the $SU(2)$ gauge group, comes exactly in three copies with respect to the electroweak interactions.

A new fourth family with mass degenerate fermions is ruled out at more than $90\%$ level of confidence since 
the associated $S$ parameter would be positive and too large. Non-degenerate heavy Dirac 
fermion doublets can substantially decrease the 
value of $S$ at the expenses of a non-zero and always positive value of the $T$ parameter. We show that for 
the present theory, a small splitting of the fermion masses is sufficient to make the model an economical 
and elegant candidate for a mechanism dynamically breaking the symmetry of the electroweak theory.
 
In figure \ref{fig:ST0} we show the results allowing for mass splitting for the leptons but keeping degenerate
 techniquarks. 
\begin{figure}[htbp]
  \begin{center}
    \mbox{
      \subfigure[Perturbative Techniquarks]{\resizebox{!}{3cm}{\includegraphics{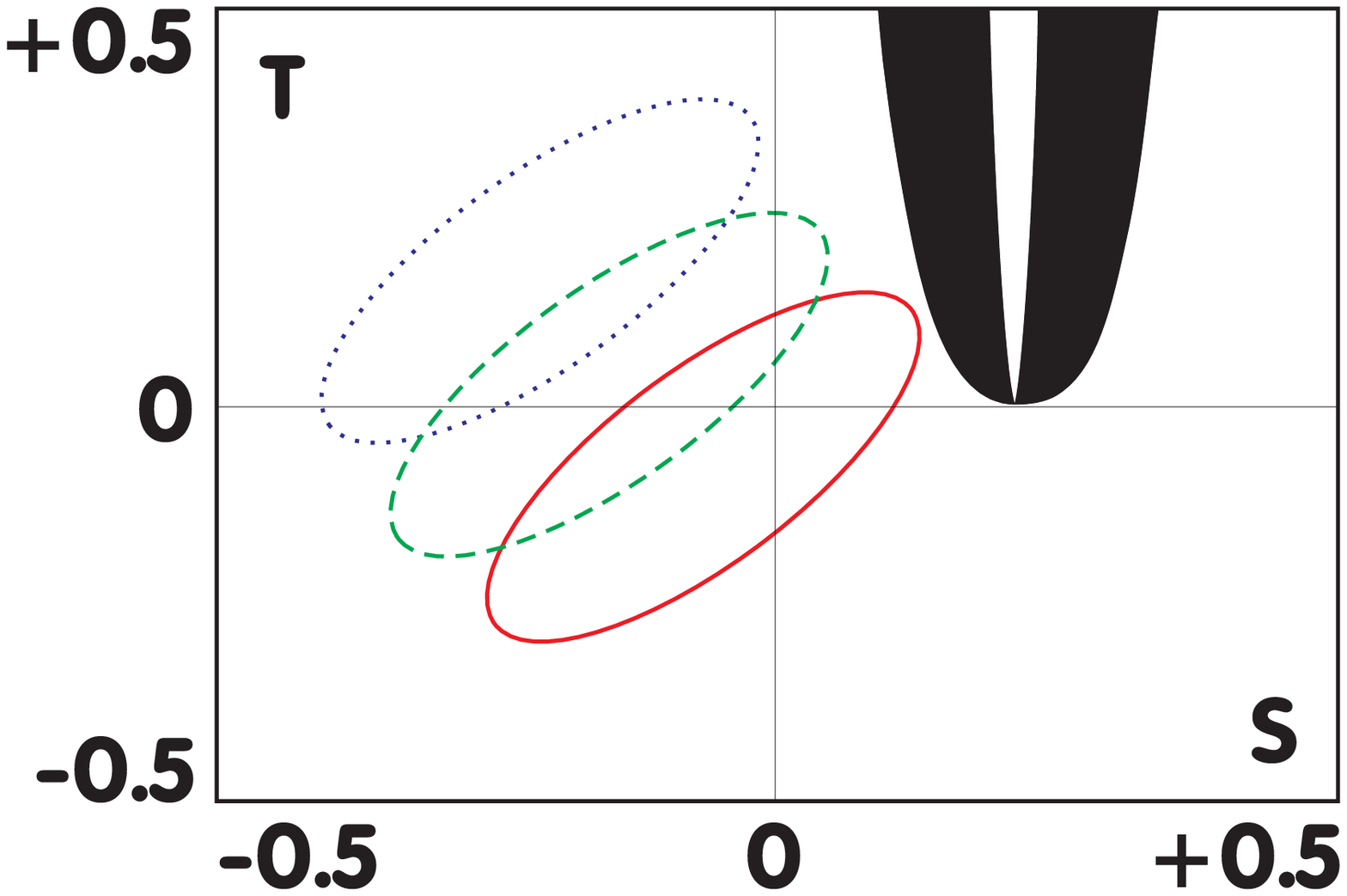}}} \qquad 
      \subfigure[Nonperturbative Techniquarks]{\resizebox{!}{3cm}{\includegraphics{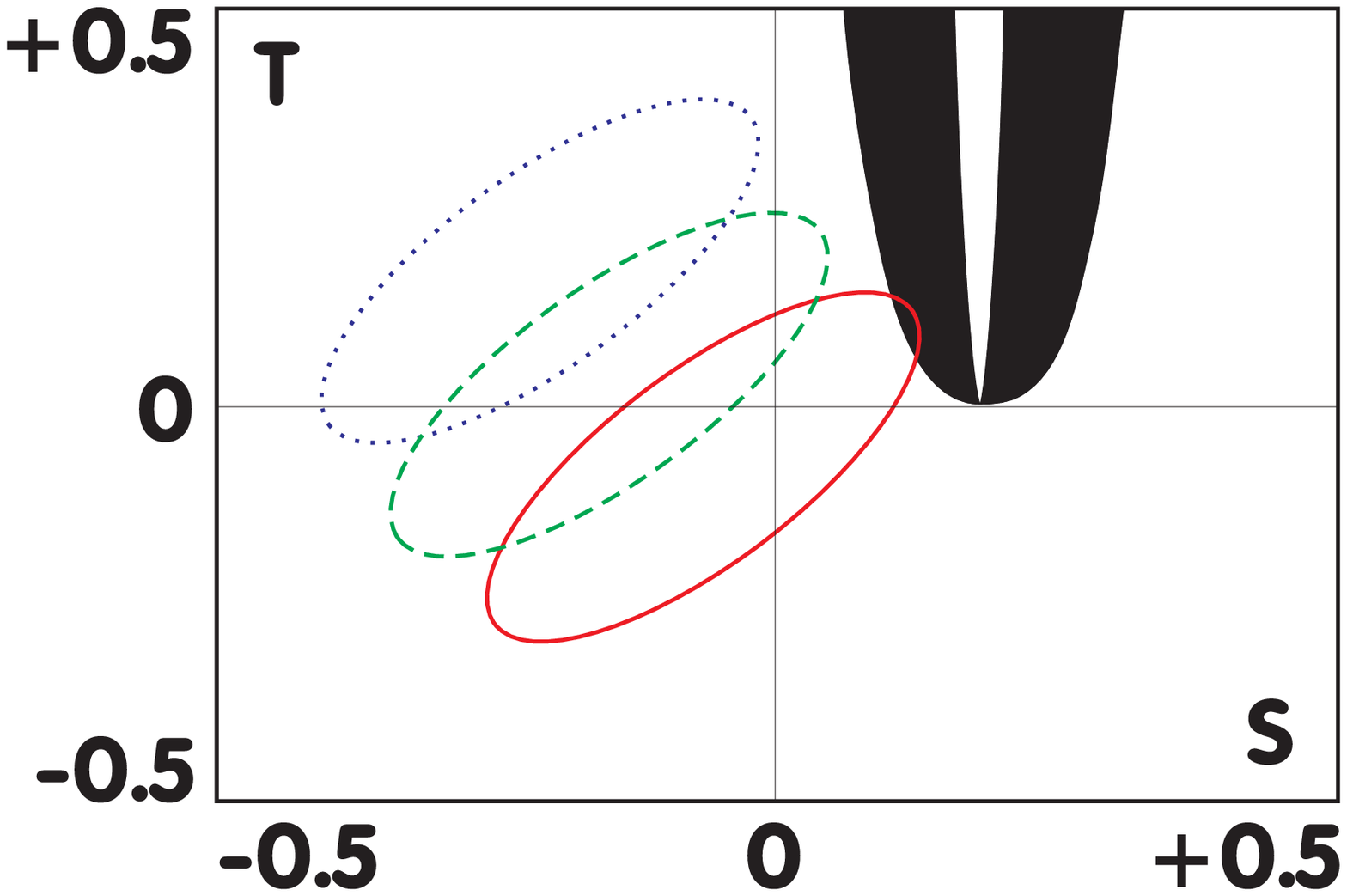}}} 
      }
    \caption{Left Panel: The black shaded parabolic area corresponds to the accessible range of $S$ and $T$ for 
    the extra neutrino and extra electron for masses from $m_Z$ to $10 m_Z$. The perturbative estimate for the 
    contribution to $S$ from techniquarks equals $1/2\pi$. The ellipses are the $90$\% confidence level contours 
for the global fit to the electroweak precision data with $U$ kept at $0$. The values of $U$ in our model lie 
typically between $0$ and $0.05$ whence they are consistent with these contours. The contours from bottom to top 
are for Higgs masses of $m_H = 117$, $340$, $1000$ GeV, respectively. Right Panel: We added non-perturbative 
corrections to the $S$ parameter in the technicolor sector of the theory.}
    \label{fig:ST0}
  \end{center}
\end{figure}
In the left panel we present the accessible range of the two oblique parameters $S$ and $T$ when the extra 
neutrino and electron masses range from $m_Z$ to $10~m_Z$. In the left panel we consider the perturbative value 
$1/2\pi$ for the contribution from the techniquarks. The ellipses are the $90\%$ confidence level contours for 
the global fit to the electroweak precision data to be found in the latest review of the Review of Particle 
Properties \cite{Eidelman:2004wy} with $U$ kept at $0$. The values of $U$ in our model lie typically between $0$ 
and $0.05$ whence they are consistent with these contours. The contours from bottom to top are for Higgs masses 
of  $m_H = 117$, $340$, $1000$ GeV, respectively. 

There are also non-perturbative corrections which further reduce the techniquark contribution to the $S$ parameter, whereby they bring the theory even 
closer to the precision measurements as can be seen from the right panel of figure \ref{fig:ST0}. 
{}To be more precise, near-conformal dynamics leads to a further reduction in
the $S$ parameter \cite{Sundrum:1991rf,Appelquist:1998xf}.
In the estimate of \cite{Sundrum:1991rf},
based
on the operator product expansion, the factor of ${1 \over 6 \pi}$
in the expression for the perturbative value of $S_{pert}=\frac{1}{6\pi} N(N+1)/2$ for one doublet of technifermions in the 
two-index symmetric representation of the $SU(N)$ gauge theory  is reduced to about $.04$, which is roughly a 
twenty percent reduction. This correction has been used to produce the figure in the right panel of figure \ref{fig:ST0}.

Note that current models of walking type with fermions in the fundamental representation are disfavored 
by the data. This is clear already when considering the perturbative as well as non-perturbative computation of the associated $S$ parameter \cite{Hong:2004td}.

We have also considered the case in which all of the new fermions have integer electric charges. In our example we have chosen the lowest possible integer charge for the fermions compatible with the absence of anomalies. The new lepton family features a doubly as well as a singly charged lepton. Interestingly, when comparing the predictions for the oblique corrections due to the model with the data we find a larger overlap with the confidence level contours (see figure \ref{fig:STY}). Such an improvement is due to the fact that the hypercharge for the new leptons is larger than in the cases investigated above, while the technicolor sector contribution is unaffected in the limit of degenerate techniquarks. 

\begin{figure}[htbp]
  \begin{center}
    \mbox{
      \subfigure[Perturbative Techniquarks]{\resizebox{!}{3cm}{\includegraphics{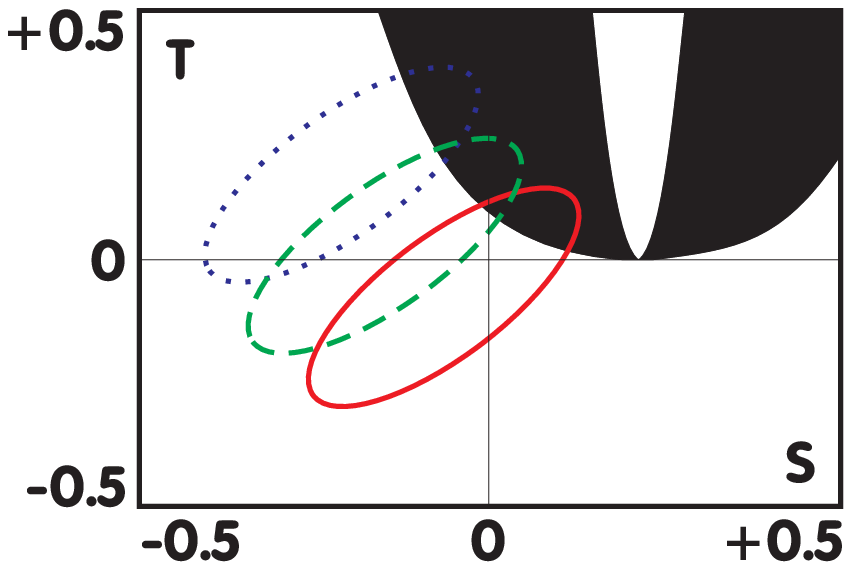}}} \qquad 
      \subfigure[Nonperturbative Techniquarks]{\resizebox{!}{3cm}{\includegraphics{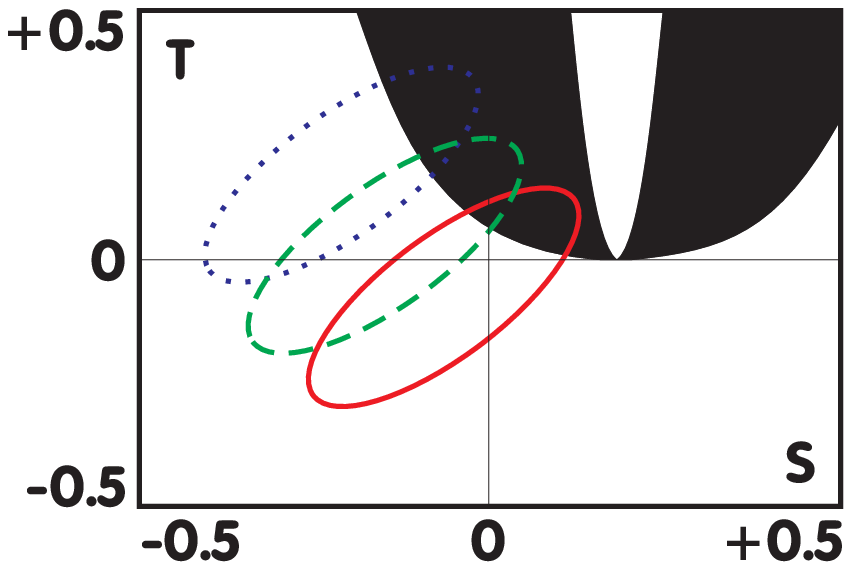}}} 
}
    \caption{Left Panel: The black shaded parabolic area corresponds to the accessible range of $S$ and $T$ for 
    the new singly and doubly charged leptons with masses from $m_Z$ to $10 m_Z$. The perturbative estimate for the 
    contribution to S from the techniquarks is $1/2\pi$. The ellipses are the $90$\% confidence level contours 
    for the global fit to the electroweak precision data with $U$ kept at $0$. The values of $U$ in our model lie 
    typically between $0$ and $0.05$ and hence they are consistent with these contours. The contours from bottom 
    to top are for Higgs masses of  $m_H = 117$, $340$, $1000$ GeV, respectively. Right Panel: We added non-perturbative corrections to the $S$ parameter in the
technicolor sector of the theory.}
    \label{fig:STY}
  \end{center}
\end{figure}

In reference \cite{Peskin:2001rw} Peskin and Wells already lined out different ways, elaborated in the past decade, to save traditional models of dynamical breaking of the electroweak sector from being ruled out by the data from precision measurements. Our results 
explicitly show that the method of positive $T$ \cite{Peskin:2001rw} is sufficient to bring the composite 
Higgs theory within the experimental acceptable range when the technifermions are in higher representations of the technicolor gauge group. 
{A heavy fourth family of ordinary quarks and/or leptons has been investigated theoretically and experimentally in the past, see \cite{Eidelman:2004wy} for an up to date review and also \cite{Froggatt:1996um,{Froggatt:2002vs}}}. 

We have also investigated the case of three technicolors and two techniflavors \cite{Dietrich:2005jn}.

\section{Lepton Spectrum at LHC and a Dark Matter candidate? }
\label{6}

The general feature of the two technicolor theory with technifermions in the two-index symmetric 
representation of the gauge group is 
the necessity to include at least one new lepton family.
If the theory underlying the spontaneous breaking of the electroweak symmetry is of the type 
presented above, the current precision measurements are already sufficient to allow us to make 
specific predictions on the new associated leptonic sector. 
This might also guide, in the future, the construction of extended technicolor models.

Assigning standard model like charges for the three 
techniquarks requires also standard model type charges for the fourth family of leptons. 
%
%
Interestingly we predict that at LHC one might discover a fourth family of ordinary leptons while the associated quarks would be bound into objects which do not interact strongly but are the technihadrons associated  to the electroweak theory. The Higgs must be light and this is consistent with our estimates provided in the next section.  

The new neutrino has a mass between $m_Z$ and $1.5~m_Z$, while the associated negatively charged lepton has a mass of roughly twice the mass of the neutral weak gauge boson. Interestingly the new neutrino, if made stable, could be a natural cold dark matter candidate. Actually this hypothesis has been investigated, in some detail, in the recent literature \cite{Belotsky:2004ga}. Here it has been found that a fourth family of neutrino is expected to be a relevant component of dark matter. Its presence might help resolving a number of yet unresolved observational puzzles \cite{Belotsky:2004ga}.

We now turn to the case of a higher hypercharge assignment featuring a singly and a doubly (electrically) charged 
lepton. In figure \ref{fig:DCL} we show the accessible range of the lepton masses. 
\begin{figure}[htbp]
  \begin{center}
    \mbox{
      \subfigure[]{\resizebox{!}{3cm}{\includegraphics{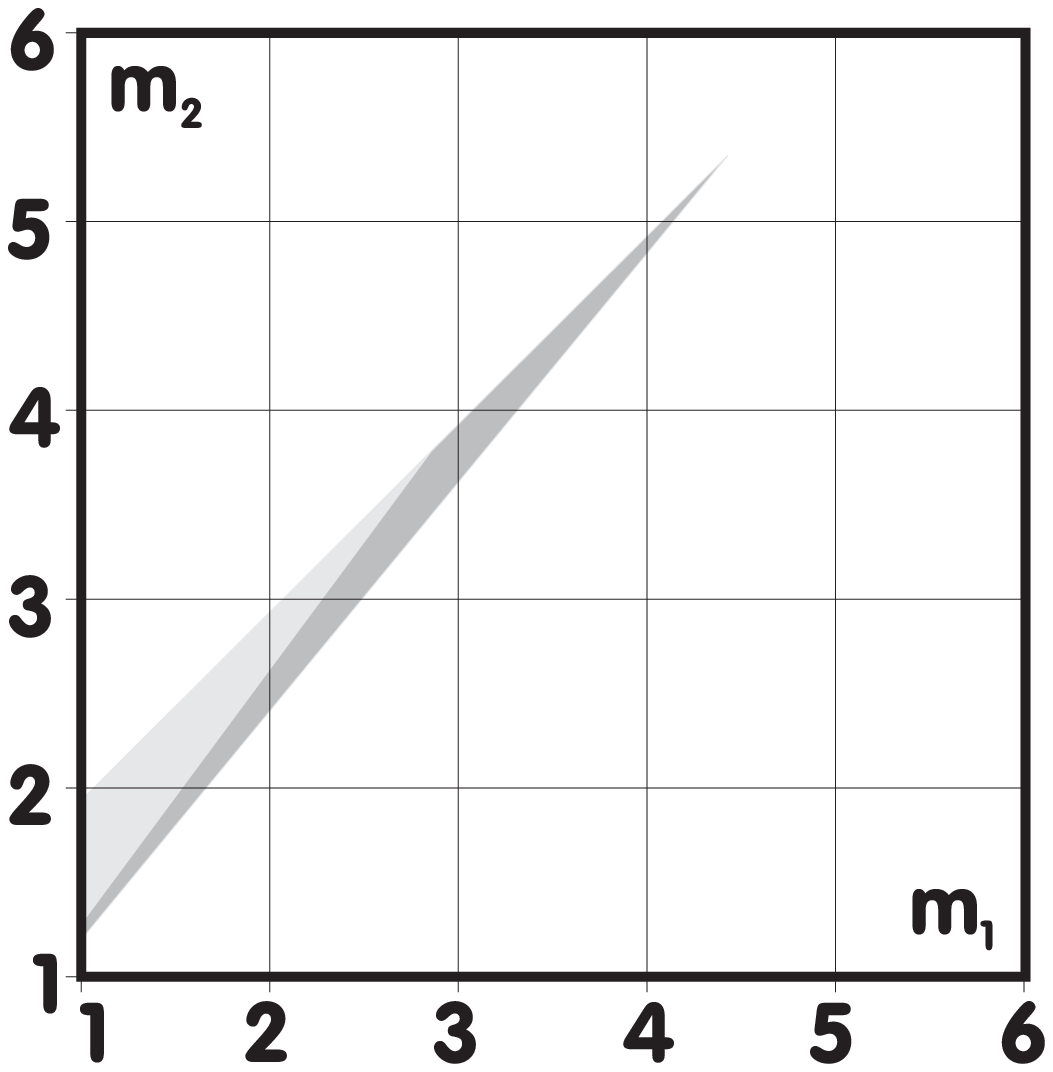}}} \qquad 
      \subfigure[]{\resizebox{!}{3cm}{\includegraphics{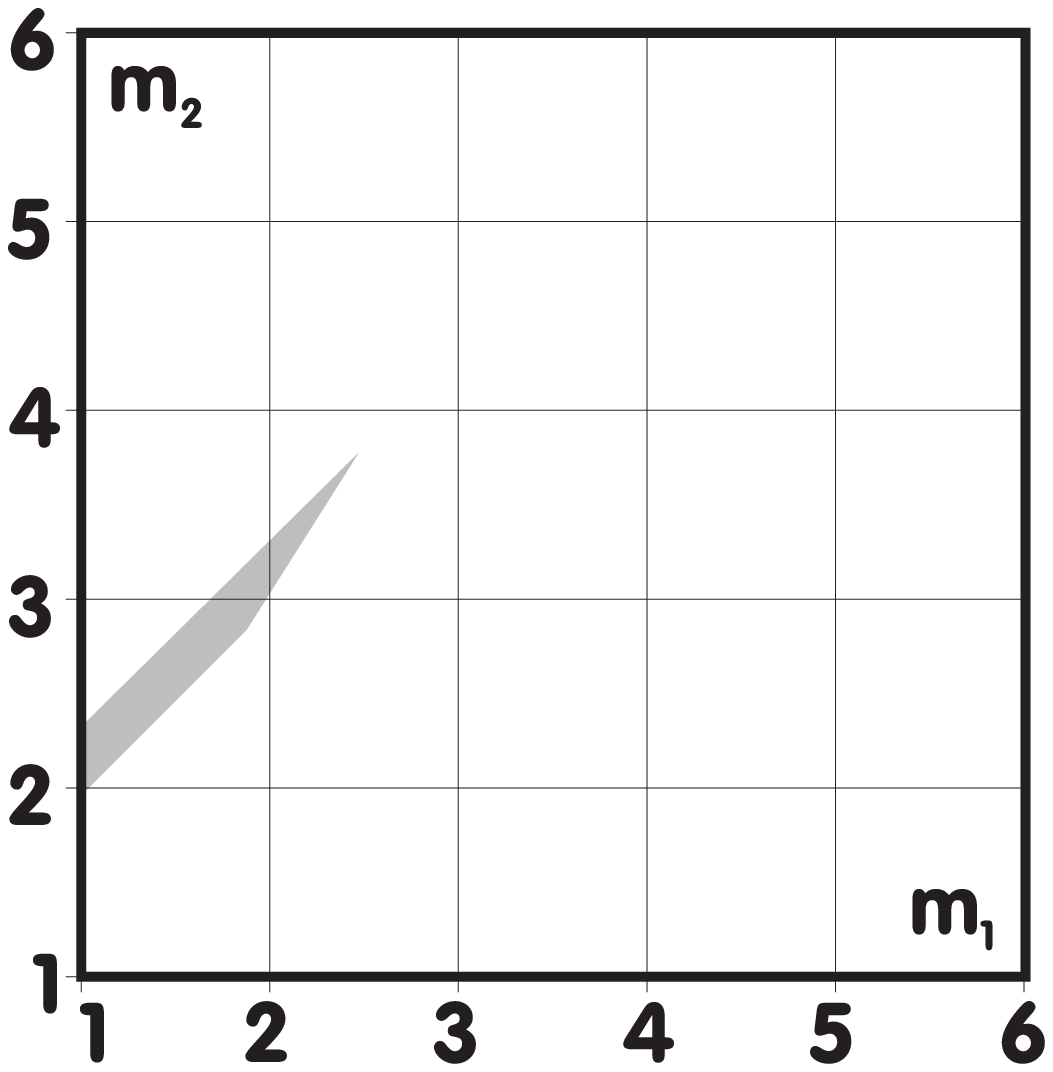}}}\qquad 
      \subfigure[]{\resizebox{!}{3cm}{\includegraphics{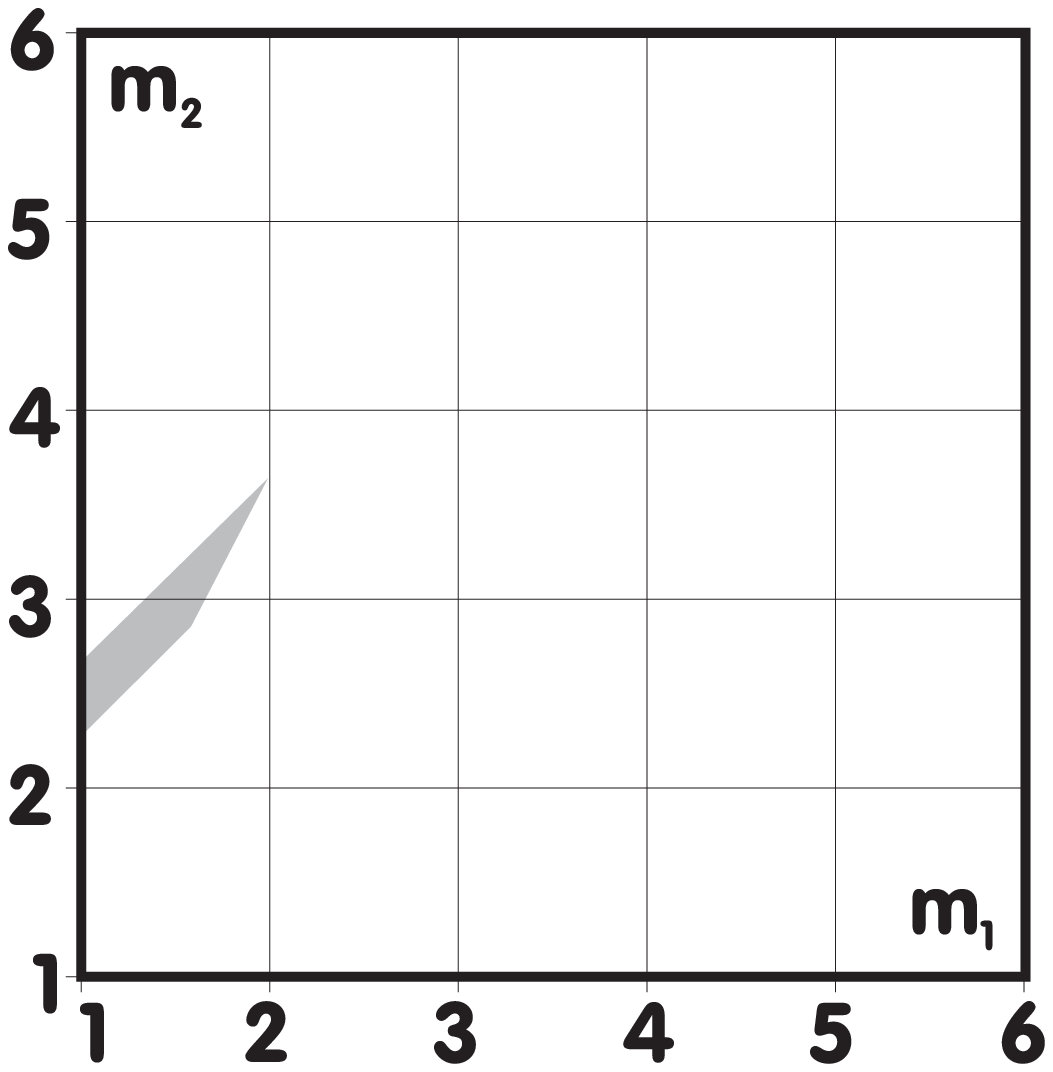}}} 
      }
    \caption{The shaded areas depict the accessible range of the new lepton masses 
    due to the oblique corrections. $m_1$($m_2$) is the mass parameter in units of $m_Z$ of the singly(doubly) charged lepton. The polygonal shapes do not correspond exactly to the ellipses in the S-T-plane but to rectangular areas defined by:
a) $S<0.12$ and $T<0.13$ for $m_H=117$~GeV;  
b) $S<0.04$ and $0.13<T<0.24$ for $m_H=340$~GeV; and c) $S<-0.02$ and $0.23<T<0.38$ for $m_H=1$~TeV. The shaded  areas in all of the figures correspond to the non-perturbative evaluation of the technicolor contribution to the $S$-parameter. In the plot for $m_H=117$~GeV the light grey part corresponds to the purely perturbative evaluation.}
    \label{fig:DCL}
  \end{center}
\end{figure}
Since we expect a very light Higgs, figure (a) is the relevant one. We then predict
 the doubly charged lepton to be heavier than the associated singly charged one. It is unstable and 
 decays into the associated singly charged lepton. The experimental bounds on the existence and properties of a doubly charged lepton are very weak \cite{Eidelman:2004wy}. 
\section{Light Composite Higgs}
\label{7}

In the analysis of QCD-like technicolor models information on the non-perturbative dynamics at the electroweak scale is obtained by simply scaling
up QCD phenomenology to the electroweak
energy scale. The Higgs particle is then mapped into the scalar chiral-partner of the Goldstone bosons in QCD. 
The scalars represent 
a very interesting and complicated sector of QCD. Much work has been devoted to providing a 
 better understanding of this sector which is relevant to understand the vacuum structure 
 of QCD. There is a growing consensus that the low lying scalar object, i.e. $f_0(600)$,  needed to provide a good description of low energy pion pion scattering \cite{Sannino:1995ik} 
 is not the chiral partner of the pions but is of four quark nature \`a la Jaffe \cite{Jaffe:1976ig,{Jaffe:1976ih}}.

Recent arguments, based on taking the limit of a large number of colors $N$, also demonstrate that the low energy scalar is not of $q\bar{q}$ nature \cite{Harada:2003em,Pelaez:2003dy,Uehara:2003ax}. The natural candidate for the chiral partner of the ordinary pions is then very heavy, i.e. it has a mass larger than one GeV, and this experimental result agrees with naive scaling estimates. When transposed to the electroweak theory by simply taking $F_{\pi}$ as the electroweak scale, one concludes that in technicolor theories with QCD-type dynamics the Higgs is very heavy, $m_H\sim 4 \pi F_{\pi}$, of the order of the TeV scale. This also means that large corrections are needed, due to new physics, to 
compensate the effects of such a heavy Higgs with respect to the electroweak precision measurements data.

{}However, for strongly interacting theories with non-QCD-like dynamics we are no longer guaranteed that the 
dynamically generated Higgs particle is heavy
\footnote{Here we are not considering theories in
which the Higgs is a quasi Goldstone boson of some strongly
interacting theory, i.e. the so called little Higgs theories.}.
In particular the QCD-like estimates cannot be
trusted in walking technicolor or other near-conformal models. This is especially true for the theories with fermions 
in higher representation of the gauge group. One
cannot simply scale up QCD to obtain useful non-pertubative
information. 

One of the main problems when considering non-QCD-like theories to construct possible extensions of the standard 
model is the lack of specific predictions on the non-perturbative dynamics. 
Very recently we have shown \cite{Hong:2004td,{Sannino:2004qp}} that it is possible 
to provide new information on the hadronic sector (and therefore the Higgs mass) of theories with higher representation by studying the one flavor sector of the S(A)-type theories. 
This was possible due to the recent observation \cite{Armoni:2004uu} that {\em
non}-supersymmetric Yang-Mills theories with a Dirac fermion
either in the two-index symmetric or antisymmetric representation
of the gauge group are non-perturbatively equivalent to
supersymmetric Yang-Mills theory (SYM) at large $N$.  At finite $N$ they were studied in \cite{Sannino:2003xe} and many of the discovered
properties, such as almost exact parity doubling and small vacuum
energy density, are appealing properties for dynamical breaking of
the electroweak theory \cite{Appelquist:1998xf}. To go beyond the one flavor case we then used the fact that 
our theories are near a phase transition \cite{Dietrich:2005jn}. 

We recall that  near the critical point of a continuous phase
transition, the mass squared of the scalar order parameter drops
proportional to $(t-t_c)^{\nu}$, with $t$ the parameter driving
the phase transition, $t_c$ its value at the transition point, and
$\nu$ the critical exponent. One well-known example is ordinary
massless QCD near the chiral symmetry restoration point at finite
temperature. In this case the scalar partner of the pions must
become light close to the phase transition. So, despite its large mass
in vacuum, the scalar meson becomes very light near the phase
transition. Continuous phase transitions asymptotically close to the critical point 
display a
classical behavior. In other words, close to the transition, the thermal
fluctuations override the quantum ones \cite{Shopova:2003jj,{Vojta},{Sachdev}}
\footnote{More precisely one has a quantum phase transition when the de Broglie thermal wavelength $\lambda$ 
is greater than the correlation length of the thermal fluctuations $\xi$, i.e. $\lambda/\xi > 1$.}. {}For zero temperature field
theory which we are discussing here, the quantum effects remain
important.

In the context of non-supersymmetric gauge theories the zero temperature chiral phase
transitions as function of the number of flavors have been studied using non-perturbative methods, e.g. in 
\cite{Appelquist:1991kn}. Here the authors already stressed that 
the phase transition as function of the number of flavors is very different from the one driven by temperature. 
Other interesting theories displaying quantum phase transitions as function of the number of flavors, although without 
gauge interactions, were investigated in \cite{Hamidian:1995pf}.

We approach the conformal window from the hadronic side of the theory by adjusting the number
of (techni)flavors with respect to the number of (techni)colors. So we treat the number of flavors $N_{Tf}$ as a
tunable scaling field. $N_{Tf}^c$ is the critical number of
flavors at which chiral symmetry is restored together with the onset of conformal invariance signaled by the vanishing of the 
underlying trace anomaly. We then construct an effective mean field theory and assume the scalar fermion-antifermion field to be the relevant one. 

We are interested in the behavior of the lightest scalar fermion-antifermion field near the phase transition and will not attempt to compute the critical exponents of the theory. The effective potential of the theory is 
\begin{eqnarray}
V[\sigma] = \frac{1}{2}M^2_{\sigma}(N_{Tf})\, \sigma^2 + \frac{\lambda}{4}\sigma^4 \ ,
\end{eqnarray}
while the kinetic term is normalized canonically. Since we are in the spontaneously 
broken phase of the theory $M^2_{\sigma}<0$ while the physical squared mass is proportional to the absolute 
value of $M^2_{\sigma}$. The associated trace of the energy momentum tensor is
\begin{eqnarray}
\theta^{\mu}_{\mu} = -M^2_{\sigma}(N_{Tf}) \sigma^2 \ .
\end{eqnarray}
Now we recall that the expression for the trace anomaly of the underlying $SU(N)$ gauge theory is
\begin{eqnarray}
\theta^{\mu}_{\mu} = -\frac{\beta}{2g} G^{\mu\nu;a}G_{\mu\nu}^a \ , \quad {\rm with} \quad {a=1,\cdots, N^2-1} \ . 
\end{eqnarray}
If the theory develops a conformal fixed point, the beta function vanishes. In order to connect the previous equation to the phase transition we are interested in we consider the 
two loop beta function 
\begin{eqnarray}\beta&=&-{\beta_0}\frac{g^3}{16\pi^2} -{\beta_1}\frac{g^5}{(16\pi^2
)^2} \ ,\quad 
\beta_0 = \frac{11}{3}N - \frac{2}{3}N_{T f}\,(N+2) \ , \nonumber \\  
\beta_1 &=&\frac{34}{3}N^2 - N_{T f} \left(N+
2\right)\left[\frac{10}{3}N + \frac{2}{N}\left(N - 
1\right)\left(N + 2\right)\right] \ . \nonumber
\end{eqnarray}
Here we have provided the coefficients for a generic $SU(N)$ S-type theory with $N_{T f}$ flavors while 
the following result is independent of the specific representation to which the fermions belong. We 
rewrite the two loop beta function as follows:
\begin{eqnarray}
-\frac{\beta}{2g}=\frac{\beta_1}{32\pi^2}\,\alpha \left(\alpha - \alpha_{\ast}\right) \ , \quad {\rm with } \quad \frac{\alpha_{\ast}}{4\pi} = -\frac{\beta_0}{\beta_1} \ .
\end{eqnarray}
As we decrease $N_{Tf}$ relative to $N$ we have that $\alpha_{\ast}$ increases. 
To extract further information we impose the extra condition that the anomalous dimension $\gamma$ of the quark operator near the phase transition assumes the value one.
This last condition is 
consistent with the fact that one expects chiral 
symmetry to be broken for $\gamma>1$ \cite{{Cohen:1988sq},{Hill:2002ap}}. Define with $\alpha_c$ the special value of the coupling constant for which the previous condition on the anomalous dimension is satisfied. We have then:
 \begin{eqnarray}
-\frac{\beta[\gamma=1]}{2g_c}=\frac{\beta_1}{32\pi^2}\,\alpha_c \left(\alpha_c - \alpha_{\ast}\right) \propto N^c_{Tf} - N_{Tf}\ .
\end{eqnarray}
{}For $\alpha_{\ast}$ above $\alpha_{c}$ chiral symmetry breaks. So, in order to compare with the effective Lagrangian theory we are  
approaching $\alpha_c$ from large values of $\alpha_{\ast}$. However the previous expression is valid on both sides of the transition. 
If the two trace anomalies, i.e. the one in the effective theory and the one in the underlying theory, are describing the same physics then 
the flavor dependence is contained in the mass of the scalar field which reads near the phase transition:
\begin{eqnarray}
|M^2_{\sigma}(N_f)|\propto (N^c_{Tf} - N_{Tf}) \ .  
\end{eqnarray}
Interestingly this is exactly the mean field theory type relation, where $N_{Tf}$ is the scaling parameter of the theory. The previous way of computing the dependence of the mass of the fermion-antifermion field on the number of flavors near the fixed point  is probably too crude of an approximation, but 
it fits well with the mean field theory approach to a phase transition. Besides, never in 
our previous approach we needed perturbation theory since we can always work in the 't Hooft scheme in which the 
two loop beta function is exact. It is only when we try to provide a value 
for the critical number of flavors that approximations must be made in non-supersymmetric theories.

We observe that the mass of the scalar is reduced from its value at $N_{Tf}=1$ as $N_f$ increases. While a great deal
is known about the phase diagram of supersymmetric gauge theories
as function of the number of flavors and colors, much less is known
about the non-supersymmetric gauge theories. 
The perturbative estimate for $N_f^c$ in the S-type theories was computed in \cite{Sannino:2004qp}.
{}For the two(three) technicolor theory it yields $N_f^c\simeq 2.1(2.5 )$. If we use
as $M_\sigma(N_{Tf}=1)=300 - 500$~GeV obtained in \cite{Hong:2004td}, we obtain
\begin{eqnarray}
m_{H}=M_\sigma(N_{Tf}=2)&\sim &90 - 150~{\rm GeV}  \ , \qquad {\rm S-two~technicolors} . 
\end{eqnarray}
Note that even if we would have chosen $1$~TeV as normalization mass for one flavor the mass of the scalar near the phase transition would still be highly suppressed. Here $M_{\sigma}$ denotes the physical mass of the scalar meson.
Interestingly, not only are we able to resolve the ordinary hierarchy problem but we can also account for a 
dynamical light Higgs. 
\begin{table}[h]
\begin{tabular}{c|c|c|c|c|c}
   & QCD-like & WTC$(3,11)$ &WTC$(2,7)$ &S$(3,2)$ & S$(2,2)$ \\
  \hline \hline
  $m_H$(GeV)~$\approx$ & $1000$ & $400$ & $300$ & $170-300$ & $90-150$ \\
  \hline \hline
\end{tabular}
\caption{Higgs mass for QCD-like theories, walking technicolor with fermions in the fundamental representation WTC$(N,N_{TF})$, and for the S-type theories S$(N,N_{Tf})$.}  
\label{tableMh}
\end{table}

{}For comparison we evaluate the mass of the scalar field (i.e. Higgs field) in the case of ordinary walking technicolor with fermions in 
the fundamental representation of the gauge group. Here we use as normalization point the case of three flavors and colors which is the 
scaled up version of QCD. In these theories the critical number of flavors is $N^c_{T f}\approx 3.9 N$ which leads to the value of the Higgs near the transition of the order of $m_{H}\approx 290$~GeV for three colors and eleven flavors. In the case of the 
two color theory with seven 
flavors while still assuming a mass of the order of one TeV for the two flavor case one obtains $m_{H} \approx 370$~GeV. In table \ref{tableMh} we summarize the expected masses for the Higgs for different types of technicolor theories.


\section{Conclusion and Outlook}
\label{8}
We have provided technicolor theories which are not ruled out by electroweak precision measurements, 
naturally yield 
a very light Higgs while predicting the existence of a new fourth lepton family with masses of the order of the 
electroweak 
scale. The Higgs is light due to a nearby quantum phase transition. We have developed a way to estimate its mass for different technicolor theories and shown that some of the theories presented here have a composite Higgs with a mass less than or equal to $150$~GeV. Possible cosmology oriented applications have been suggested. 
{It is tempting to speculate that the cosmological constant problem  may be resolved by assuming a nearby 
quantum phase transition for the universe. In this scenario the cosmological constant is expected to be the function 
of a yet unknown scaling parameter, similar to the number of flavors for the Higgs mass, which turns out to be 
very close to its critical value.}

While we have assumed a bottom-up approach and constructed explicit technicolor theories which are not ruled out by 
the current precision measurements, it is also very important, at this point, to consider extensions capable of 
addressing the mass generation problem in some detail. 

The unification of gauge couplings has not been addressed here but will be investigated in the near future.
{}For this purpose, it is then interesting to observe that the two technicolor theory can 
be rewritten as an $SO(3)$ theory with fermions in the fundamental representation of the gauge group which is a theory similar to the one investigated in \cite{Giudice:1991sz}.

The finite temperature electroweak phase transition within the present theories is also an interesting avenue to 
explore. The outcome is relevant for the baryogenesis problem since the baryon asymmetry cannot be explained 
within the standard model. We point out that the presence of 
higher-order operators in the Higgs field are naturally expected to emerge when breaking the 
electroweak theory using the technicolor theories presented here. The cut-off in the Higgs Lagrangian seen as a low energy effective theory, is naturally identified with the mass of the excited hadronic states which have been integrated out at low energy. The resulting 
effective theory would emerge in a fashion very similar to the toy model presented in \cite{Grojean:2004xa}.

\section*{Acknowledgments}

I would like to thank A.H. Fariborz for organizing the MRST meeting 2005 and for creating such an interesting scientific environment. It is my pleasure to thank D.D. Dietrich, D.K. Hong, S.D. Hsu 
 and K. Tuominen for having shared the work on which this proceeding is based. 
The work is supported in part by the Marie Curie Excellence Grant as team leader under contract MEXT-CT-2004-013510 and the Skou fellowship of the Danish Research Agency


\begin{thebibliography}{000}    
\bibitem{TC}
S.~Weinberg,
Phys.\ Rev.\ D {\bf 19}, 1277 (1979);
L.~Susskind,
Phys.\ Rev.\ D {\bf 20}, 2619 (1979).

\bibitem{Hill:2002ap}
C.~T.~Hill and E.~H.~Simmons,
`
Phys.\ Rept.\ {\bf 381}, 235 (2003) [Erratum-ibid.\ {\bf 390},
553 (2004)].

\bibitem{Lane:2002wv}
  K.~Lane,
  arXiv:hep-ph/0202255.

\bibitem{Peskin:2001rw}
M.~E.~Peskin and J.~D.~Wells,
Phys.\ Rev.\ D {\bf 64}, 093003 (2001)
[arXiv:hep-ph/0101342].



\bibitem{Peskin:1991sw}
  M.~E.~Peskin and T.~Takeuchi,
  %
  Phys.\ Rev.\ D {\bf 46}, 381 (1992).



\bibitem{Sannino:2004qp}
F.~Sannino and K.~Tuominen,
Phys.\ Rev.\ D {\bf 71}, 051901 (2005).
arXiv:hep-ph/0405209.

\bibitem{Hong:2004td}
  D.~K.~Hong, S.~D.~H.~Hsu and F.~Sannino,
  Phys.\ Lett.\ B {\bf 597}, 89 (2004)
  [arXiv:hep-ph/0406200].



\bibitem{Dietrich:2005jn}
  D.~D.~Dietrich, F.~Sannino and K.~Tuominen,
  arXiv:hep-ph/0505059.




\bibitem{Dimopoulos:1981xc}
  S.~Dimopoulos and J.~Preskill,
  Nucl.\ Phys.\ B {\bf 199}, 206 (1982).
  D.~B.~Kaplan and H.~Georgi,
  Phys.\ Lett.\ B {\bf 136}, 183 (1984).




\bibitem{Arkani-Hamed:2001nc}
  N.~Arkani-Hamed, A.~G.~Cohen and H.~Georgi,
  Phys.\ Lett.\ B {\bf 513}, 232 (2001)
  [arXiv:hep-ph/0105239].



\bibitem{Holdom:1981rm}
B.~Holdom,
Phys.\ Rev.\ D {\bf 24}, 1441 (1981).

\bibitem{Yamawaki:1985zg}
K.~Yamawaki, M.~Bando and K.~i.~Matumoto,
Phys.\ Rev.\ Lett.\ {\bf 56}, 1335 (1986).

\bibitem{Appelquist:an}
T.~W.~Appelquist, D.~Karabali and L.~C.~R.~Wijewardhana,
Phys.\ Rev.\ Lett.\ {\bf 57}, 957 (1986).

\bibitem{MY}
  V.~A.~Miransky and K.~Yamawaki,
  Phys.\ Rev.\ D {\bf 55}, 5051 (1997)
  [Erratum-ibid.\ D {\bf 56}, 3768 (1997)]
  [arXiv:hep-th/9611142].
  V.~A.~Miransky, T.~Nonoyama and K.~Yamawaki,
  Mod.\ Phys.\ Lett.\ A {\bf 4}, 1409 (1989).


\bibitem{Frere:1986ct}
  J.~M.~Frere,
  Phys.\ Rev.\ D {\bf 35}, 2625 (1987).

\bibitem{Lane:1989ej}
  K.~D.~Lane and E.~Eichten,
  Phys.\ Lett.\ B {\bf 222}, 274 (1989).

\bibitem{Lane:1991qh}
  K.~D.~Lane and M.~V.~Ramana,
  Phys.\ Rev.\ D {\bf 44}, 2678 (1991).

\bibitem{Corrigan:1979xf}
  E.~Corrigan and P.~Ramond,
  Phys.\ Lett.\ B {\bf 87}, 73 (1979).


\bibitem{Intriligator:1995au}
  K.~A.~Intriligator and N.~Seiberg,
  Nucl.\ Phys.\ Proc.\ Suppl.\  {\bf 45BC}, 1 (1996)
  [arXiv:hep-th/9509066].


\bibitem{Hsu:2004mf}
  S.~D.~H.~Hsu and F.~Sannino,
  Phys.\ Lett.\ B {\bf 605}, 369 (2005)
  [arXiv:hep-ph/0408319].


\cite{Arkani-Hamed:2004fb}
\bibitem{Arkani-Hamed:2004fb}
  N.~Arkani-Hamed and S.~Dimopoulos,
  arXiv:hep-th/0405159.

\bibitem{Cohen:1988sq}
  A.~G.~Cohen and H.~Georgi,
  Nucl.\ Phys.\ B {\bf 314}, 7 (1989).


\bibitem{Appelquist:2002me}
  T.~Appelquist and R.~Shrock,
  breaking,''
  Phys.\ Lett.\ B {\bf 548}, 204 (2002)
  [arXiv:hep-ph/0204141].

\bibitem{Appelquist:2003uu}
  T.~Appelquist and R.~Shrock,
  Phys.\ Rev.\ Lett.\  {\bf 90}, 201801 (2003)
  [arXiv:hep-ph/0301108].

  T.~Appelquist, M.~Piai and R.~Shrock,
  Phys.\ Rev.\ D {\bf 69}, 015002 (2004)
  [arXiv:hep-ph/0308061].


\bibitem{Appelquist:2004es}
  T.~Appelquist, M.~Piai and R.~Shrock,
  Phys.\ Lett.\ B {\bf 595}, 442 (2004)
  [arXiv:hep-ph/0406032].

\bibitem{Appelquist:2004ai}
  T.~Appelquist, N.~Christensen, M.~Piai and R.~Shrock,
  Phys.\ Rev.\ D {\bf 70}, 093010 (2004)
  [arXiv:hep-ph/0409035].

\bibitem{Appelquist:2004mn}
  T.~Appelquist, M.~Piai and R.~Shrock,
  Phys.\ Lett.\ B {\bf 593}, 175 (2004)
  [arXiv:hep-ph/0401114].


\bibitem{Witten:fp}
E.~Witten,
Phys.\ Lett.\ B {\bf 117}, 324 (1982).




\bibitem{Ringwald:2004np}
  A.~Ringwald and Y.~Y.~Y.~Wong,
  JCAP {\bf 0412}, 005 (2004)
  [arXiv:hep-ph/0408241].




\bibitem{reusser91}
D.~Reusser {\it et al.}, 
Phys.\ Lett.\ B {\bf 255} (1991) 143. 

\bibitem{Abusaidi:2000wg}
  R.~Abusaidi {\it et al.}  [CDMS Collaboration],
  Phys.\ Rev.\ Lett.\  {\bf 84}, 5699 (2000)
  [arXiv:astro-ph/0002471].



\bibitem{Chivukula:1989qb}
R.~S.~Chivukula and T.~P.~Walker,
Nucl.\ Phys.\ B {\bf 329}, 445 (1990).
J.~Bagnasco, M.~Dine and S.~Thomas,
Phys.\ Lett.\ B {\bf 320}, 99 (1994)
[arXiv:hep-ph/9310290].

\bibitem{Peskin:1990zt}
  M.~E.~Peskin and T.~Takeuchi,
  %
  Phys.\ Rev.\ Lett.\  {\bf 65}, 964 (1990).


\bibitem{Kennedy:1990ib}
  D.~C.~Kennedy and P.~Langacker,
  %
  Phys.\ Rev.\ Lett.\  {\bf 65}, 2967 (1990)
  [Erratum-ibid.\  {\bf 66}, 395 (1991)].

\bibitem{Altarelli:1990zd}
  G.~Altarelli and R.~Barbieri,
  %
  Phys.\ Lett.\ B {\bf 253}, 161 (1991).
  



\bibitem{He:2001tp}
  H.~J.~He, N.~Polonsky and S.~f.~Su,
  Phys.\ Rev.\ D {\bf 64}, 053004 (2001)
  [arXiv:hep-ph/0102144].






\bibitem{Eidelman:2004wy}
  S.~Eidelman {\it et al.}  [Particle Data Group],
  ``Review of particle physics,''
  Phys.\ Lett.\ B {\bf 592}, 1 (2004).
  
\bibitem{Sundrum:1991rf}
  R.~Sundrum and S.~D.~H.~Hsu,
  Nucl.\ Phys.\ B {\bf 391}, 127 (1993)
  [arXiv:hep-ph/9206225].

\bibitem{Appelquist:1998xf}
T.~Appelquist and F.~Sannino,
Phys.\ Rev.\ D {\bf 59}, 067702 (1999) [arXiv:hep-ph/9806409].
T.~Appelquist, P.~S.~Rodrigues da Silva and F.~Sannino,
Phys.\ Rev.\ D {\bf 60}, 116007 (1999) [arXiv:hep-ph/9906555].
Z.~y.~Duan, P.~S.~Rodrigues da Silva and F.~Sannino,
Nucl.\ Phys.\ B {\bf 592}, 371 (2001) [arXiv:hep-ph/0001303].


\bibitem{Froggatt:1996um}
  C.~D.~Froggatt, D.~J.~Smith and H.~B.~Nielsen,
  Z.\ Phys.\ C {\bf 73}, 333 (1997)
  [arXiv:hep-ph/9603436].


\bibitem{Froggatt:2002vs}
  C.~D.~Froggatt and J.~E.~Dubicki,
  arXiv:hep-ph/0203146.

\bibitem{Belotsky:2004ga}
  K.~Belotsky, D.~Fargion, M.~Y.~Khlopov, R.~V.~Konoplich, M.~G.~Ryskin and K.~I.~Shibaev,
  arXiv:hep-ph/0411271.
  K.~Belotsky, D.~Fargion, M.~Khlopov and R.~V.~Konoplich,
  arXiv:hep-ph/0411093.
  K.~Belotsky, D.~Fargion, M.~Khlopov, R.~Konoplich and K.~Shibaev,
  Phys.\ Rev.\ D {\bf 68}, 054027 (2003)
  [arXiv:hep-ph/0210153].
  D.~Fargion, Y.~A.~Golubkov, M.~Y.~Khlopov, R.~V.~Konoplich and R.~Mignani,
  Pisma Zh.\ Eksp.\ Teor.\ Fiz.\  {\bf 69}, 402 (1999)
  [JETP Lett.\  {\bf 69}, 434 (1999)]
  [arXiv:astro-ph/9903086].








\bibitem{Sannino:1995ik}
F.~Sannino and J.~Schechter,
Phys.\ Rev.\ D {\bf 52}, 96 (1995) [arXiv:hep-ph/9501417];
M.~Harada, F.~Sannino and J.~Schechter,
Phys.\ Rev.\ D {\bf 54}, 1991 (1996) [arXiv:hep-ph/9511335];
M.~Harada, F.~Sannino and J.~Schechter,
Phys.\ Rev.\ Lett.\  {\bf 78}, 1603 (1997) [arXiv:hep-ph/9609428];
D.~Black, A.~H.~Fariborz, F.~Sannino and J.~Schechter,
Phys.\ Rev.\ D {\bf 58}, 054012 (1998) [arXiv:hep-ph/9804273];
D.~Black, A.~H.~Fariborz, F.~Sannino and J.~Schechter,
Phys.\ Rev.\ D {\bf 59}, 074026 (1999) [arXiv:hep-ph/9808415].

\bibitem{Jaffe:1976ig}
  R.~L.~Jaffe,
   Mesons,''
  %
  Phys.\ Rev.\ D {\bf 15}, 267 (1977).

\bibitem{Jaffe:1976ih}
  R.~L.~Jaffe,
  %
  Phys.\ Rev.\ D {\bf 15}, 281 (1977).

\bibitem{Harada:2003em}
M.~Harada, F.~Sannino and J.~Schechter,
Phys.\ Rev.\ D {\bf 69}, 034005 (2004) [arXiv:hep-ph/0309206].

\bibitem{Pelaez:2003dy}
  J.~R.~Pelaez,
  Phys.\ Rev.\ Lett.\  {\bf 92}, 102001 (2004)
  [arXiv:hep-ph/0309292].


\bibitem{Uehara:2003ax}
  M.~Uehara,
  %
  arXiv:hep-ph/0308241.





\bibitem{Armoni:2004uu}
A.~Armoni, M.~Shifman and G.~Veneziano,
arXiv:hep-th/0403071;
  A.~Armoni, M.~Shifman and G.~Veneziano,
  Phys.\ Rev.\ D {\bf 71}, 045015 (2005)
  [arXiv:hep-th/0412203].
  
\bibitem{Sannino:2003xe}
  F.~Sannino and M.~Shifman,
  Phys.\ Rev.\ D {\bf 69}, 125004 (2004)
  [arXiv:hep-th/0309252].

  
  
  
\bibitem{Shopova:2003jj}
  D.~V.~Shopova and D.~I.~Uzunov,
  ``Some basic aspects of quantum phase transitions,''
  Phys.\ Rept.\  {\bf 379}, 1 (2003)
  [arXiv:cond-mat/0301423].


\bibitem{Vojta}
 M.~ Vojta, ``Quantum Phase Transitions'', Rep. Prog. Phys. 66, 2069 (2003) 
 [arXiv:cond-mat/0309604].

\bibitem{Sachdev}
S.~Sachdev, ``Quantum Phase Transitions'', Cambridge University Press, Cambridge, 1999.



\bibitem{Appelquist:1991kn}
  T.~Appelquist, J.~Terning and L.~C.~R.~Wijewardhana,
  Phys.\ Rev.\ D {\bf 44}, 871 (1991);
  T.~Appelquist, J.~Terning and L.~C.~R.~Wijewardhana,
  Phys.\ Rev.\ Lett.\  {\bf 77}, 1214 (1996)
  [arXiv:hep-ph/9602385];
  T.~Appelquist, A.~Ratnaweera, J.~Terning and L.~C.~R.~Wijewardhana,
  Phys.\ Rev.\ D {\bf 58}, 105017 (1998)
  [arXiv:hep-ph/9806472].

\bibitem{Hamidian:1995pf}
  H.~Hamidian, G.~W.~Semenoff, P.~Suranyi and L.~C.~R.~Wijewardhana,
  Phys.\ Rev.\ Lett.\  {\bf 74}, 4976 (1995)
  [arXiv:hep-ph/9502303].
  
  
\bibitem{Giudice:1991sz}
  G.~F.~Giudice and S.~Raby,
  Nucl.\ Phys.\ B {\bf 368}, 221 (1992).
  
\bibitem{Grojean:2004xa}
  C.~Grojean, G.~Servant and J.~D.~Wells,
  Phys.\ Rev.\ D {\bf 71}, 036001 (2005)
  [arXiv:hep-ph/0407019]. See also references therein.




\end{thebibliography}
\end{document}